\documentclass[aps,prb,twocolumn,showpacs,superscriptaddress]{revtex4}
\usepackage{graphicx}
\usepackage[latin1]{inputenc}
\usepackage{dcolumn}
\usepackage{bm}

\begin{document}

\title{Terahertz-induced depletion of the ground-state population of neutral donors in GaAs measured by resonant elastic light scattering from donor-bound excitons}

\author{D. G. Allen}
\affiliation{Department of Physics, University of California,
Santa Barbara, California 93106} \email{sherwin@physics.ucsb.edu}
\author{C. R. Stanley}
\affiliation{Department of Electronics and Electrical Engineering,
University of Glasgow, Glasgow, G12 8QQ, UK}
\author{M. S. Sherwin}
\affiliation{Department of Physics, University of California,
Santa Barbara, California 93106} \email{sherwin@physics.ucsb.edu}
\date{\today}

\begin{abstract}
Strong resonant elastic light scattering (RELS) from the
donor-bound exciton transition in GaAs (1.514eV) occurs at neutral
donors in the ground (1S) state, but not at neutral donors in
excited hydrogenic states. When 1.6 THz radiation is incident on
an ensemble of neutral donors, we observe up to a 30\% decrease in
the RELS, corresponding to a decrease in the population of neutral
donors in their ground states. This optical detection method is
similar to quantum nondemolition measurement techniques used for
readout of ion trap quantum computers and diamond nitrogen-vacancy
centers. In this scheme, Auger recombination of the bound exciton,
which changes the state of the donor during measurement, limits
the measurement fidelity and maximum NIR excitation intensity.
\end{abstract}


\maketitle

Neutral shallow donors in semiconductors provide a model system
for the study of quantum information in solid state materials for
both spin-based\cite{kane:nature,vrijenQC} and electronic
orbital-based qubits.\cite{colenature} Comprising a positively
charged donor and a single bound electron, neutral shallow donors
exhibit a hydrogen-like energy spectrum scaled by the electron
effective mass ratio ($m^\ast/m_e$) and the inverse square of the
relative dielectric constant ($\epsilon_r$). In hydrogenic
orbital-based neutral donor qubits, the 1S and 2P levels of the
bound electron serve as qubit states. For GaAs, the binding energy
of electrons to hydrogenic donors, such as S or Si, is
$\sim$5.9meV, resulting bound state transition frequencies
$\sim$1THz. These are well below the LO phonon energy (36meV,
8.7THz), leading to long excited state lifetimes
($t_1=350$ns).\cite{2plifetime,2pminlifetime} Cole, et al.
demonstrated Rabi oscillations in ensembles of shallow donors in
GaAs, using photoconductivity as the measure of the excited state
population.\cite{colenature} However, photoconductivity is only
indirectly sensitive to the state of the donor\cite{dotythesis}
and completely destroys the qubit. It is desirable to have an
alternative means of readout which: (1) is directly sensitive to
the state of the bound electron, (2) allows rapid distinguishing
between states of the qubit, and (3) achieves a quantum
nondemolition (QND) measurement (i.e. leaves the qubit in the
state reported by the measurement).

In ion trap quantum computers\cite{CiracZoller} QND measurement is
achieved by using an optical probe. The qubit states are hyperfine
states of the trapped ion, and readout is done by resonantly
exciting atoms from the lowest hyperfine state to an auxiliary
state. Selection rules prevent a spontaneous relaxation from the
auxiliary state to other hyperfine states. During readout, atoms
in the lowest hyperfine state continually absorb and re-emit, or
scatter, the incident light and are visible, whereas atoms in
excited hyperfine states remain dark.\cite{wineland:primer} A
photoluminescence-based optical readout technique has been used to
observe magnetic resonance\cite{diamNVnmr} and coherent quantum
dynamics in single nitrogen-vacancy
centers.\cite{diamNVsinglespin} In this report, we introduce the
use of resonant near infrared (NIR) excitation of the neutral
donor ($\textrm{D}^0$) ground (1S) state to donor-bound exciton
($\textrm{D}^0$X) transition as a means of determining whether
neutral donors are in the ground (1S) state. Absorption and
re-emission, i.e. scattering, of the NIR light occurs at donors in
the 1S state, whereas donors in excited hydrogenic states do not
interact with the light. Using this technique we observe changes
in the population of donors in the 1S state due to excitation of
the donor-bound electron with THz radiation.

Detection of visible/NIR emission as a means of observing THz/FIR
dynamics in semiconductors is a well-used technique. In related
work, FIR (THz) modulated photoluminescence (PL) has been used to
observe phenomena in a variety of semiconductor structures,
including bulk GaAs,\cite{stanleyODCR} undoped quantum
wells,\cite{motr} coupled quantum wells,\cite{firmpl} and InGaAs
quantum dots.\cite{InGaAsQDdblres} McCombe and coworkers have used
a variety of optical detection schemes for FIR spectroscopy
including both nonresonant (PL) and resonant (specularly
reflected) emission.\cite{odr} Our technique differs from the
above optically detected FIR studies in that we measure scattered,
nonspecular emission at the NIR excitation frequency, rather than
PL only, or specular reflection.

Shallow or hydrogenic donors have been the subject of considerable
theoretical and experimental work.\cite{impurityreview} In
particular, the $\textrm{D}^0$X state has received a great deal of
attention. Karasyuk et al.\ utilized $\textrm{D}^0$X transitions
to identify donors in GaAs as well as determine donor binding
energies and central cell corrections.\cite{D0X} (Measurements of
GaAs neutral donor properties have also been made by FIR
absorption spectroscopy and photoconductivity.\cite{D0Xabsspec})

\begin{figure}
\includegraphics[width=2.5in]{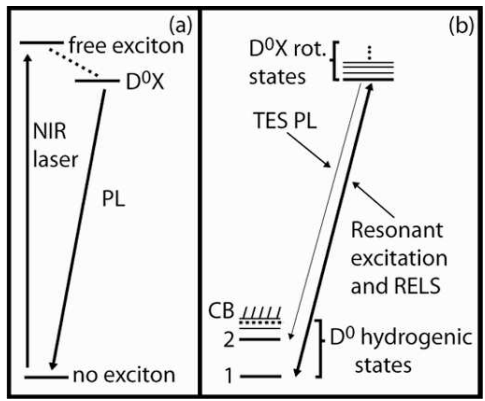}
\caption{(a)  Photoluminescence (PL) and resonant elastic light
scattering (RELS) from donor-bound excitons ($\textrm{D}^0$X). (a)
A NIR laser creates a free exciton which loses energy via phonon
emission, is captured by a neutral donor, and emits PL. (b)  A NIR
laser is tuned to a $\textrm{D}^0$X resonance.  Both RELS and PL
occur. In both (a) and (b), decay to the $\textrm{D}^0$1S state
via direct exciton recombination is fast ($<1$ns). Weak PL with a
lower frequency results from exciton decay with simultaneous
excitation of the donor-bound electron to a higher hydrogenic
orbital (Auger recombination).\label{diagram}}
 \end{figure}

\begin{figure}\includegraphics{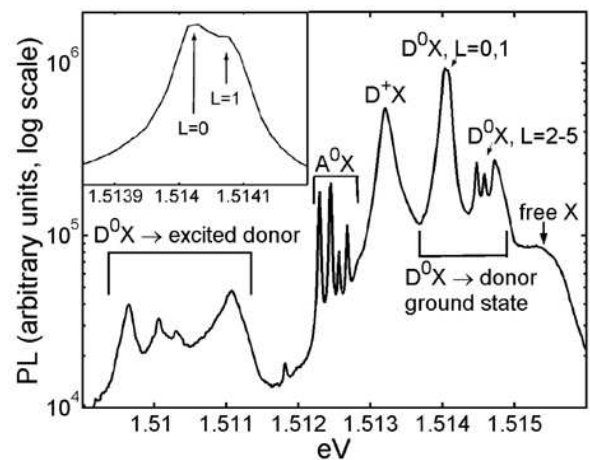}\caption{5K PL spectrum
(log scale) of GaAs ($N_d-N_a=3\times 10^{14}\textrm{cm}^{-3}$)
taken with CCD and grating spectrometer at $35\mu$eV (0.02nm)
resolution. Laser excitation was at 1.52eV, $1\textrm{mWcm}^{-2}$
intensity. Labels identify transitions associated with the free
exciton ($free \ X$), the neutral donor-bound exciton
($\textrm{D}^0$X), the ionized donor-bound exciton
($\textrm{D}^+$X), and the neutral acceptor-bound exciton
($\textrm{A}^0$X). $L$ denotes the rotational quantum number of
the $\textrm{D}^0$X initial state. (Inset) Linear scale PL of
direct exciton recombination from the two lowest $\textrm{D}^0$X
rotational levels.\label{pl}}
\end{figure}

The donor-bound exciton state ($\textrm{D}^0$X) is a short-lived
$H_2$-like complex (binding energy $\sim$1meV) with several
rotational bound states. The bound exciton recombines optically
with an average lifetime of 1ns (232ps in a quantum
well).\cite{D0Xoscstrength,quantumwellD0X} Donor-bound excitons
may be created by above band gap excitation of free excitons
followed by capture via phonon emission (Fig. \ref{diagram}a). In
that case, the light emitted from bound exciton recombination is
PL (see Fig. \ref{pl}). Alternatively, the sample may be excited
below the band gap, at the $\textrm{D}^0$X resonance (Fig.
\ref{diagram}b). Light emitted from the sample at the laser
frequency may be due to either resonant Rayleigh scattering or
resonance fluorescence (light absorbed and re-emitted by
$\textrm{D}^0$X recombination). We use the term resonant elastic
light scattering (RELS) to refer to all the light collected at the
laser frequency.\cite{donorRRS}$^,$\footnote{Although authors in
Ref. \onlinecite{donorRRS} observed polarization-preserving
resonant Rayleigh scattering from the $\textrm{D}^0$X transition,
under the current experimental conditions we were unable to
measure an increase in co-polarized scattered light on resonance,
so the resonant emission we observed is primarily due to resonance
fluorescence.} Since the probability of decay from a
$\textrm{D}^0$X state to the neutral donor 1S state
($\textrm{D}^0$1S) is high, donors measured to be in the 1S state
continually cycle between the 1S state and the $\textrm{D}^0$X
state, elastically scattering many NIR photons. Donors in excited
hydrogenic states do not absorb the incident light, and are dark.

When the bound exciton decays, there is a nontrivial probability
of transferring energy to the electron bound to the donor, leaving
it in an excited state. If that occurs, the state of the donor is
changed during the measurement. This Auger recombination process
is the limiting factor in the ability of the $\textrm{D}^0$X RELS
technique to make a QND measurement of a neutral donor. Although
$\textrm{D}^0$X RELS does not yet provide a high fidelity QND
measurement, the probability of return to the ground state is high
enough to enable sensitive detection of population electrons in
the D01S state, under low intensity NIR excitation conditions.

\begin{figure}\includegraphics{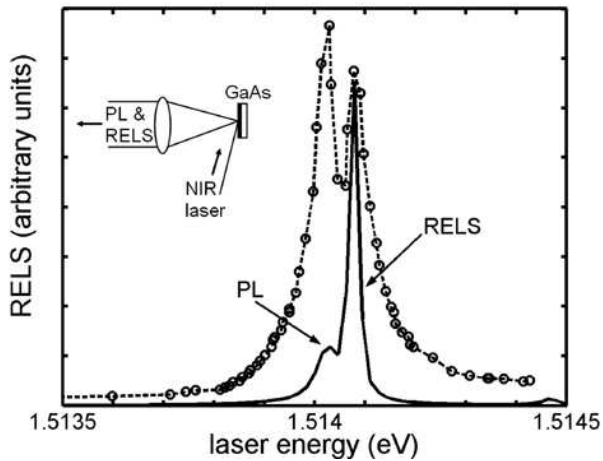}
\caption{5K GaAs emission spectrum (solid line) showing RELS at
the laser excitation energy and PL. Open circles trace the RELS
peak height versus excitation energy. The dashed line is a guide
to the eye. (Inset) Experimental geometry for collection of NIR
emission. \label{resfl}}\end{figure}

Samples for all experiments consisted of a 15$\mu$m layer of
unintentionally-doped high-purity GaAs grown by molecular beam
epitaxy on a $500\mu$m thick semi-insulating GaAs substrate. The
effective donor density ($N_d-N_a$) was inferred to be $3\times
10^{14}\textrm{cm}^{-3}$ from Hall effect measurements. FIR
photoconductivity indicates the dominant donor impurities are S
and Si, and the PL spectrum confirms acceptors are a relatively
minor impurity. For optical measurements, a GaAs sample was
mounted in a strain-free manner on the cold finger of a liquid
helium flow cryostat with a minimum sample temperature near 5K.
The sample was excited in the NIR by a tunable external cavity
diode laser with typical intensities $\sim 1\textrm{mWcm}^{-2}$,
incident at Brewster's angle. Emission, including both PL and
elastically scattered light, was collected from the surface normal
by a lens and focused into a 0.75m imaging spectrometer and
detected by a CCD or PMT (see inset to Fig. \ref{resfl}). A $CO_2$
laser-pumped difluoromethane gas laser was used for FIR (THz)
excitation. Characteristic incident intensities were less than
$20\textrm{mWcm}^{-2}$.

Fig. \ref{pl} shows a typical 5K GaAs PL spectrum. Peak
assignments follow those of Ref. \onlinecite{D0X}. Transitions
from various $\textrm{D}^0$X rotational states to the $D^0$1S
state (direct recombination) occur in the region 1.5139-1.515eV.
Transitions involving Auger recombination, often called two
electron satellite (TES) transitions in semiconductor literature,
leave the donor in an excited state. These are visible in the
region 1.509-1.5115eV. The observed recombination of excitons
bound to ionized donors ($\textrm{D}^+\textrm{X}$) near 1.513eV
occurs primarily in the $\sim 1\mu$m depletion region near the
unpassivated surface. The PL associated with excitons bound to
neutral acceptors ($\textrm{A}^0\textrm{X}$) occurs between 1.512
and 1.513eV.

The solid line in Fig. \ref{resfl} is a typical emission spectrum
with excitation in the region of the two lowest direct
$\textrm{D}^0$X transitions (labelled L=0 and L=1 in Fig. \ref{pl}
inset). RELS is much stronger than PL, and strongest when
excitation is resonant with a $\textrm{D}^0$X transition. Open
circles in Fig. \ref{resfl} trace the RELS peak height from a
series of emission spectra taken across a range excitation
energies. The two inhomogeneously broadened transitions are much
better resolved by RELS than by PL (compare Fig. \ref{resfl} with
inset to Fig \ref{pl}).

\begin{figure}
\includegraphics{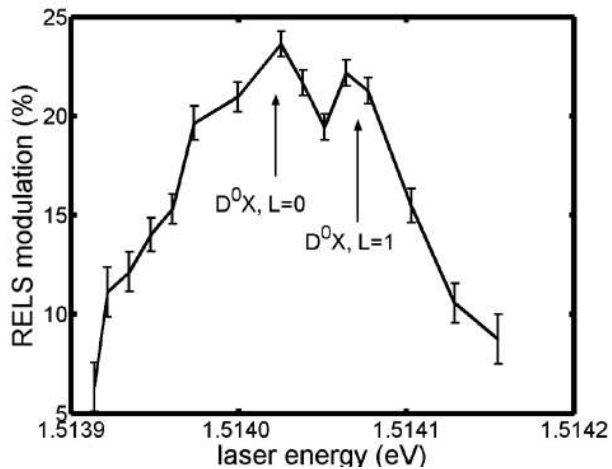}
\caption{RELS modulation versus incident NIR laser frequency
(intensity $\sim 1\textrm{mWcm}^{-2}$). Sample temperature is 5K
and the incident FIR power is $\sim 20\textrm{mWcm}^{-2}$. The
error bars are due to fluctuations in the intensity of the FIR
laser. Labels denote the spectral features shown in Fig.
\ref{pl}(inset) and Fig. \ref{resfl}.\label{mod}}
\end{figure}

When FIR photons with enough energy to promote bound electrons to
a higher hydrogenic state, or the conduction band, are incident on
the sample, fewer donors are in their ground state, and hence less
RELS from $\textrm{D}^0$X to $\textrm{D}^0$1S transitions is
observed. Fig. \ref{mod}
shows the percent change in the RELS at several NIR excitation
energies with FIR light of frequency 1.63THz (6.73meV) and
$20\textrm{mWcm}^{-2}$ intensity incident on the sample. Since the
FIR excitation energy is greater than the 5.9meV binding energy of
the donors, electrons are excited directly to the conduction band.
The percent change in RELS, or RELS modulation, is calculated as
$100(A-B)/[(A+B)/2]$, where $B$ and $A$ are the respective RELS
signals with and without FIR excitation. The curve in Fig.
\ref{mod} indicates the modulation is greatest when the NIR is
resonant with a $\textrm{D}^0$X transition. The experimentally
observed modulation depends on the relative sizes and overlap of
the NIR and FIR laser spots, the NIR laser frequency and
intensity, FIR laser intensity and sample temperature. The maximum
experimentally observed value of the modulation with FIR
excitation at 1.6THz was 30\%. Modulation of the RELS due to FIR
excitation at 1.4THz (5.78meV, below the electron ionization
energy) or 1.04THz (4.31meV, slightly detuned from the 1S-2P bound
transition) is similar in magnitude. In future experiments, a
freely tunable FIR source such as a free electron laser or energy
level tuning via magnetic field could be used to investigate RELS
modulation under conditions where the FIR is in resonance with the
1S-2P transition, in order to observe Rabi oscillations, or other
resonant or coherent effects.

\begin{figure}
\includegraphics{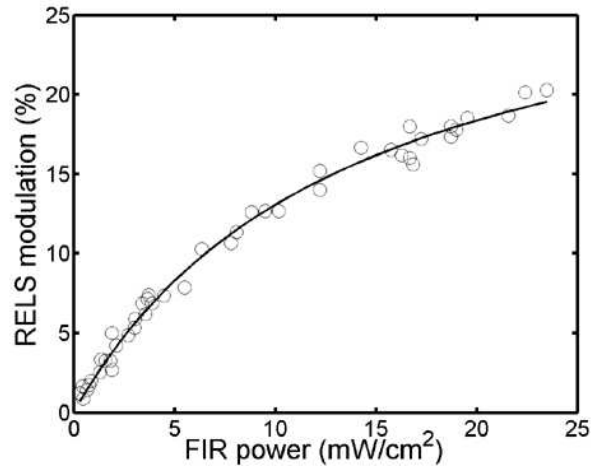}
\caption{RELS modulation versus incident FIR intensity. Open
circles are data points. The solid line is a fit to a the curve
$A[1-(1+I/I_0)^{-1}]$. The fit parameters are $I_0=13.7$ and
$A=31$.\label{power}}
\end{figure}

RELS efficiency, like photoconductivity, is sensitive to changes
in sample temperature.  RELS efficiency drops (approximately
linearly) to 1/7 of the 5K value by 15K, and then more slowly to
$1/10$ of the 5K value by 20K. (This behavior is expected, because
the binding energy of excitons to the neutral donors is only 1meV,
and $k_BT=1\textrm{meV}$ at 11.6K.) In order to be useful for
qubit readout, the measured change in RELS would have to be due to
a change in the donor ground state population, and not merely a
reduction in RE intensity due to lattice heating. Fig. \ref{power}
 shows the dependence of the RELS
modulation on incident FIR intensity. The modulation initially
saturates at $30\%$, a value much lower than is achievable by
thermal effects ($B=1/7$ or $90\%$). The fact that the change in
RELS is constant to within 1\% for FIR modulation frequencies
$\leq 400$Hz, implies the thermal time constant ($\tau$) of the
sample is $<1/10$kHz. The corresponding temperature rise ($\Delta
T$) in terms of the absorbed power ($P_{abs}$), mass ($m$) and
specific heat ($c_p$) is approximately given by $\Delta
T=P_{abs}\tau/(m c_p)=0.2K$. Also, the linear response at low THz
intensities, where thermal effects are smallest, is consistent
with absorption modulation. Therefore, we are confident the
modulation is dominated by changes in the population of donors in
their ground states, not heating.

For qubit readout, an important figure of merit related to
measurement fidelity is the ratio of the amount of light emitted
from direct $\textrm{D}^0$X to $\textrm{D}^0\textrm{1S}$
transitions to the amount of light collected from state-altering
Auger transitions. Exciting at 1.5142eV (above the two lowest
direct $\textrm{D}^0$X transitions), the amount of light collected
in the direct transition region (1.5138-1.515eV) is 133 times
larger than light collected in the TES region (1.509-1.5117eV).
Exciting on resonance with the $\textrm{D}^0$X, L=0 transition at
1.514eV, the ratio increases to a maximum of 1250, while the
integrated emission in the direct transition region increases by a
factor of 60. This indicates resonant excitation is superior to PL
and/or off resonant excitation from both a signal detection and
measurement fidelity standpoint.

Measurement of donor ground state occupation via $\textrm{D}^0$X
RELS may provide a solution for single qubit readout, provided
adjacent donors can be optically resolved, and Auger transitions
can be reduced to a manageable level via, for instance, choice of
magnetic field, crystal orientation and NIR polarization. The
$\textrm{D}^0$X lifetime is 1ns, so a single donor may scatter up
to $10^9$ photons per second, if excited near saturation, and no
Auger transitions to excited states occur. For a collection
efficiency of 0.1, $10$ photons could be collected in 100ns (10
photons per 23ns in a quantum well). Currently, the collected RELS
is limited by the NIR excitation intensity, which must be kept
below $1\textrm{mWcm}^{-2}$. Higher NIR intensities result in
decreased RELS modulation (-3dB at $4\pm 1\textrm{mWcm}^{-2}$).
The reason for the decrease in RELS modulation with increasing NIR
intensity remains to be investigated, but could be due to donors
becoming shelved in long-lived excited hydrogenic states via Auger
transitions, or increased scattering from free excitons.

We note that $\textrm{D}^0$X RELS provides an alternative means of
detection for studying properties of bound electron quantum
states, in addition to absorption saturation spectroscopy and
photoconductivity. One such application may be time resolving the
recovery of donors to the ground state to determine bound-to-bound
state lifetimes.

The authors gratefully acknowledge funding support from
DARPA/QUIST, CNID (Center for Nanoscience Innovation for Defense),
and Sun Microsystems.

\bibliography{dansbib}

\begin{thebibliography}{21}
\expandafter\ifx\csname natexlab\endcsname\relax\def\natexlab#1{#1}\fi
\expandafter\ifx\csname bibnamefont\endcsname\relax
  \def\bibnamefont#1{#1}\fi
\expandafter\ifx\csname bibfnamefont\endcsname\relax
  \def\bibfnamefont#1{#1}\fi
\expandafter\ifx\csname citenamefont\endcsname\relax
  \def\citenamefont#1{#1}\fi
\expandafter\ifx\csname url\endcsname\relax
  \def\url#1{\texttt{#1}}\fi
\expandafter\ifx\csname urlprefix\endcsname\relax\def\urlprefix{URL }\fi
\providecommand{\bibinfo}[2]{#2}
\providecommand{\eprint}[2][]{\url{#2}}

\bibitem[{\citenamefont{Kane}(1998)}]{kane:nature}
\bibinfo{author}{\bibfnamefont{B.~E.} \bibnamefont{Kane}},
  \bibinfo{journal}{Nature} \textbf{\bibinfo{volume}{383}},
  \bibinfo{pages}{133} (\bibinfo{year}{1998}).

\bibitem[{\citenamefont{Vrijen et~al.}(2000)\citenamefont{Vrijen, Yablonovitch,
  Wang, Jiang, Balandin, Roychowdhury, Mor, and DiVincenzo}}]{vrijenQC}
\bibinfo{author}{\bibfnamefont{R.}~\bibnamefont{Vrijen}},
  \bibinfo{author}{\bibfnamefont{E.}~\bibnamefont{Yablonovitch}},
  \bibinfo{author}{\bibfnamefont{K.}~\bibnamefont{Wang}},
  \bibinfo{author}{\bibfnamefont{H.~W.} \bibnamefont{Jiang}},
  \bibinfo{author}{\bibfnamefont{A.}~\bibnamefont{Balandin}},
  \bibinfo{author}{\bibfnamefont{V.}~\bibnamefont{Roychowdhury}},
  \bibinfo{author}{\bibfnamefont{T.}~\bibnamefont{Mor}}, \bibnamefont{and}
  \bibinfo{author}{\bibfnamefont{D.}~\bibnamefont{DiVincenzo}},
  \bibinfo{journal}{Phys. Rev. A.} \textbf{\bibinfo{volume}{62}},
  \bibinfo{pages}{12306} (\bibinfo{year}{2000}).

\bibitem[{\citenamefont{Cole et~al.}(2001)\citenamefont{Cole, Williams, King,
  Sherwin, and Stanley}}]{colenature}
\bibinfo{author}{\bibfnamefont{B.~E.} \bibnamefont{Cole}},
  \bibinfo{author}{\bibfnamefont{J.~B.} \bibnamefont{Williams}},
  \bibinfo{author}{\bibfnamefont{B.~T.} \bibnamefont{King}},
  \bibinfo{author}{\bibfnamefont{M.~S.} \bibnamefont{Sherwin}},
  \bibnamefont{and} \bibinfo{author}{\bibfnamefont{C.~R.}
  \bibnamefont{Stanley}}, \bibinfo{journal}{Nature}
  \textbf{\bibinfo{volume}{410}}, \bibinfo{pages}{60} (\bibinfo{year}{2001}).

\bibitem[{\citenamefont{Burghoon et~al.}(1994)\citenamefont{Burghoon, Klaassen,
  and Wenchebach}}]{2plifetime}
\bibinfo{author}{\bibfnamefont{J.}~\bibnamefont{Burghoon}},
  \bibinfo{author}{\bibfnamefont{T.~O.} \bibnamefont{Klaassen}},
  \bibnamefont{and} \bibinfo{author}{\bibfnamefont{W.~T.}
  \bibnamefont{Wenchebach}}, \bibinfo{journal}{Semicond. Sci. Tech.}
  \textbf{\bibinfo{volume}{9}}, \bibinfo{pages}{30} (\bibinfo{year}{1994}).

\bibitem[{\citenamefont{Kalkman et~al.}(1996)\citenamefont{Kalkman, Pellemans,
  Klaassen, and Wencheback}}]{2pminlifetime}
\bibinfo{author}{\bibfnamefont{A.~J.} \bibnamefont{Kalkman}},
  \bibinfo{author}{\bibfnamefont{H.~P.~M.} \bibnamefont{Pellemans}},
  \bibinfo{author}{\bibfnamefont{T.~O.} \bibnamefont{Klaassen}},
  \bibnamefont{and} \bibinfo{author}{\bibfnamefont{W.~T.}
  \bibnamefont{Wencheback}}, \bibinfo{journal}{Int. J. Infrared Milli.}
  \textbf{\bibinfo{volume}{17}}, \bibinfo{pages}{569} (\bibinfo{year}{1996}).

\bibitem[{\citenamefont{Doty}(2004)}]{dotythesis}
\bibinfo{author}{\bibfnamefont{M.}~\bibnamefont{Doty}}, Ph.D. thesis,
  \bibinfo{school}{University of California, Santa Barbara}
  (\bibinfo{year}{2004}).

\bibitem[{\citenamefont{Cirac and Zoller}(1995)}]{CiracZoller}
\bibinfo{author}{\bibfnamefont{J.~I.} \bibnamefont{Cirac}} \bibnamefont{and}
  \bibinfo{author}{\bibfnamefont{P.}~\bibnamefont{Zoller}},
  \bibinfo{journal}{Phys. Rev. Lett.} \textbf{\bibinfo{volume}{74}},
  \bibinfo{pages}{4091} (\bibinfo{year}{1995}).

\bibitem[{\citenamefont{Wineland et~al.}(1998)\citenamefont{Wineland, Monreo,
  Itano, King, Liebfried, Meekhof, Myatt, and Wood}}]{wineland:primer}
\bibinfo{author}{\bibfnamefont{D.~J.} \bibnamefont{Wineland}},
  \bibinfo{author}{\bibfnamefont{C.~R.} \bibnamefont{Monreo}},
  \bibinfo{author}{\bibfnamefont{W.~M.} \bibnamefont{Itano}},
  \bibinfo{author}{\bibfnamefont{B.~E.} \bibnamefont{King}},
  \bibinfo{author}{\bibfnamefont{D.}~\bibnamefont{Liebfried}},
  \bibinfo{author}{\bibfnamefont{D.~M.} \bibnamefont{Meekhof}},
  \bibinfo{author}{\bibfnamefont{C.~J.} \bibnamefont{Myatt}}, \bibnamefont{and}
  \bibinfo{author}{\bibfnamefont{C.~S.} \bibnamefont{Wood}},
  \bibinfo{journal}{Fortschritte de Physik} \textbf{\bibinfo{volume}{46}},
  \bibinfo{pages}{363} (\bibinfo{year}{1998}).

\bibitem[{\citenamefont{Gruber et~al.}(1997)\citenamefont{Gruber,
  Dr\"abenstedt, Tietz, Fleury, Wrachtrup, and von Borczyskowski}}]{diamNVnmr}
\bibinfo{author}{\bibfnamefont{A.}~\bibnamefont{Gruber}},
  \bibinfo{author}{\bibfnamefont{A.}~\bibnamefont{Dr\"abenstedt}},
  \bibinfo{author}{\bibfnamefont{C.}~\bibnamefont{Tietz}},
  \bibinfo{author}{\bibfnamefont{L.}~\bibnamefont{Fleury}},
  \bibinfo{author}{\bibfnamefont{J.}~\bibnamefont{Wrachtrup}},
  \bibnamefont{and} \bibinfo{author}{\bibfnamefont{C.}~\bibnamefont{von
  Borczyskowski}}, \bibinfo{journal}{Science} \textbf{\bibinfo{volume}{276}},
  \bibinfo{pages}{2012} (\bibinfo{year}{1997}).

\bibitem[{\citenamefont{Jelezko et~al.}(2004)\citenamefont{Jelezko, Gaebel,
  Popa, Gruber, and Wrachtrup}}]{diamNVsinglespin}
\bibinfo{author}{\bibfnamefont{F.}~\bibnamefont{Jelezko}},
  \bibinfo{author}{\bibfnamefont{T.}~\bibnamefont{Gaebel}},
  \bibinfo{author}{\bibfnamefont{I.}~\bibnamefont{Popa}},
  \bibinfo{author}{\bibfnamefont{A.}~\bibnamefont{Gruber}}, \bibnamefont{and}
  \bibinfo{author}{\bibfnamefont{J.}~\bibnamefont{Wrachtrup}},
  \bibinfo{journal}{Phys. Rev. Lett.} \textbf{\bibinfo{volume}{92}},
  \bibinfo{pages}{076401} (\bibinfo{year}{2004}).

\bibitem[{\citenamefont{Michels et~al.}(1994)\citenamefont{Michels, Warburton,
  Nicholas, and Stanley}}]{stanleyODCR}
\bibinfo{author}{\bibfnamefont{J.~G.} \bibnamefont{Michels}},
  \bibinfo{author}{\bibfnamefont{R.~J.} \bibnamefont{Warburton}},
  \bibinfo{author}{\bibfnamefont{R.~J.} \bibnamefont{Nicholas}},
  \bibnamefont{and} \bibinfo{author}{\bibfnamefont{C.~R.}
  \bibnamefont{Stanley}}, \bibinfo{journal}{Semicond. Sci. Technol.}
  \textbf{\bibinfo{volume}{9}}, \bibinfo{pages}{198} (\bibinfo{year}{1994}).

\bibitem[{\citenamefont{Cerne et~al.}(2002)\citenamefont{Cerne, Kono, Su, and
  Sherwin}}]{motr}
\bibinfo{author}{\bibfnamefont{J.}~\bibnamefont{Cerne}},
  \bibinfo{author}{\bibfnamefont{J.}~\bibnamefont{Kono}},
  \bibinfo{author}{\bibfnamefont{M.}~\bibnamefont{Su}}, \bibnamefont{and}
  \bibinfo{author}{\bibfnamefont{M.~S.} \bibnamefont{Sherwin}},
  \bibinfo{journal}{Phys. Rev. B} \textbf{\bibinfo{volume}{66}},
  \bibinfo{pages}{205301} (\bibinfo{year}{2002}).

\bibitem[{\citenamefont{Stone et~al.}(1997)\citenamefont{Stone, Michels,
  Kinder, Chang, Nichols, Fox, and Roberts}}]{firmpl}
\bibinfo{author}{\bibfnamefont{R.~J.} \bibnamefont{Stone}},
  \bibinfo{author}{\bibfnamefont{J.~G.} \bibnamefont{Michels}},
  \bibinfo{author}{\bibfnamefont{D.}~\bibnamefont{Kinder}},
  \bibinfo{author}{\bibfnamefont{C.~C.} \bibnamefont{Chang}},
  \bibinfo{author}{\bibfnamefont{R.~J.} \bibnamefont{Nichols}},
  \bibinfo{author}{\bibfnamefont{A.~M.} \bibnamefont{Fox}}, \bibnamefont{and}
  \bibinfo{author}{\bibfnamefont{J.~S.} \bibnamefont{Roberts}},
  \bibinfo{journal}{Superlattices Microst.} \textbf{\bibinfo{volume}{21}},
  \bibinfo{pages}{597} (\bibinfo{year}{1997}).

\bibitem[{\citenamefont{Murdin et~al.}(2000)\citenamefont{Murdin, Hollingworth,
  Barker, Clarke, Findlay, Pidgeon, Wells, Bradley, Malik, and
  Murray}}]{InGaAsQDdblres}
\bibinfo{author}{\bibfnamefont{B.~N.} \bibnamefont{Murdin}},
  \bibinfo{author}{\bibfnamefont{A.~R.} \bibnamefont{Hollingworth}},
  \bibinfo{author}{\bibfnamefont{J.~A.} \bibnamefont{Barker}},
  \bibinfo{author}{\bibfnamefont{D.~G.} \bibnamefont{Clarke}},
  \bibinfo{author}{\bibfnamefont{P.~C.} \bibnamefont{Findlay}},
  \bibinfo{author}{\bibfnamefont{C.~R.} \bibnamefont{Pidgeon}},
  \bibinfo{author}{\bibfnamefont{J.-P.~R.} \bibnamefont{Wells}},
  \bibinfo{author}{\bibfnamefont{I.~V.} \bibnamefont{Bradley}},
  \bibinfo{author}{\bibfnamefont{S.}~\bibnamefont{Malik}}, \bibnamefont{and}
  \bibinfo{author}{\bibfnamefont{R.}~\bibnamefont{Murray}},
  \bibinfo{journal}{Phys. Rev. B} \textbf{\bibinfo{volume}{62}},
  \bibinfo{pages}{R7755} (\bibinfo{year}{2000}).

\bibitem[{\citenamefont{Kioseoglou et~al.}(2000)\citenamefont{Kioseoglou,
  Cheong, Yeo, Nickel, Petrou, McCombe, Sivachenko, Dzyubenko, and
  Schaff}}]{odr}
\bibinfo{author}{\bibfnamefont{G.}~\bibnamefont{Kioseoglou}},
  \bibinfo{author}{\bibfnamefont{H.~D.} \bibnamefont{Cheong}},
  \bibinfo{author}{\bibfnamefont{T.}~\bibnamefont{Yeo}},
  \bibinfo{author}{\bibfnamefont{H.~A.} \bibnamefont{Nickel}},
  \bibinfo{author}{\bibfnamefont{A.}~\bibnamefont{Petrou}},
  \bibinfo{author}{\bibfnamefont{B.~D.} \bibnamefont{McCombe}},
  \bibinfo{author}{\bibfnamefont{A.~Y.} \bibnamefont{Sivachenko}},
  \bibinfo{author}{\bibfnamefont{A.~B.} \bibnamefont{Dzyubenko}},
  \bibnamefont{and} \bibinfo{author}{\bibfnamefont{W.}~\bibnamefont{Schaff}},
  \bibinfo{journal}{Phys. Rev. B} \textbf{\bibinfo{volume}{61}},
  \bibinfo{pages}{5556} (\bibinfo{year}{2000}).

\bibitem[{\citenamefont{Pantelides}(1978)}]{impurityreview}
\bibinfo{author}{\bibfnamefont{S.~T.} \bibnamefont{Pantelides}},
  \bibinfo{journal}{Rev. Mod. Phys.} \textbf{\bibinfo{volume}{50}},
  \bibinfo{pages}{797} (\bibinfo{year}{1978}).

\bibitem[{\citenamefont{Karasyuk et~al.}(1994)\citenamefont{Karasyuk, Beckett,
  Nissen, Villemaire, Steiner, and Thewalt}}]{D0X}
\bibinfo{author}{\bibfnamefont{V.~A.} \bibnamefont{Karasyuk}},
  \bibinfo{author}{\bibfnamefont{D.~G.~S.} \bibnamefont{Beckett}},
  \bibinfo{author}{\bibfnamefont{M.~K.} \bibnamefont{Nissen}},
  \bibinfo{author}{\bibfnamefont{A.}~\bibnamefont{Villemaire}},
  \bibinfo{author}{\bibfnamefont{T.~W.} \bibnamefont{Steiner}},
  \bibnamefont{and} \bibinfo{author}{\bibfnamefont{M.~L.~W.}
  \bibnamefont{Thewalt}}, \bibinfo{journal}{Phys. Rev. B}
  \textbf{\bibinfo{volume}{49}}, \bibinfo{pages}{16381} (\bibinfo{year}{1994}).

\bibitem[{\citenamefont{Planken et~al.}(1995)\citenamefont{Planken, van Son,
  Hovenier, Klaassen, Wenckebach, Murdin, and Knippels}}]{D0Xabsspec}
\bibinfo{author}{\bibfnamefont{P.~C.~M.} \bibnamefont{Planken}},
  \bibinfo{author}{\bibfnamefont{P.~C.} \bibnamefont{van Son}},
  \bibinfo{author}{\bibfnamefont{J.~N.} \bibnamefont{Hovenier}},
  \bibinfo{author}{\bibfnamefont{T.~O.} \bibnamefont{Klaassen}},
  \bibinfo{author}{\bibfnamefont{W.~T.} \bibnamefont{Wenckebach}},
  \bibinfo{author}{\bibfnamefont{B.~M.} \bibnamefont{Murdin}},
  \bibnamefont{and} \bibinfo{author}{\bibfnamefont{G.~M.~H.}
  \bibnamefont{Knippels}}, \bibinfo{journal}{Phys. Rev. B}
  \textbf{\bibinfo{volume}{51}}, \bibinfo{pages}{9643} (\bibinfo{year}{1995}).

\bibitem[{\citenamefont{Finkman et~al.}(1986)\citenamefont{Finkman, Sturge, and
  Bhat}}]{D0Xoscstrength}
\bibinfo{author}{\bibfnamefont{E.}~\bibnamefont{Finkman}},
  \bibinfo{author}{\bibfnamefont{M.~D.} \bibnamefont{Sturge}},
  \bibnamefont{and} \bibinfo{author}{\bibfnamefont{R.}~\bibnamefont{Bhat}},
  \bibinfo{journal}{J. Lumin.} \textbf{\bibinfo{volume}{35}},
  \bibinfo{pages}{235} (\bibinfo{year}{1986}).

\bibitem[{\citenamefont{Charbonneau et~al.}(1988)\citenamefont{Charbonneau,
  Steiner, Thewalt, Koteles, Chi, and Elman}}]{quantumwellD0X}
\bibinfo{author}{\bibfnamefont{S.}~\bibnamefont{Charbonneau}},
  \bibinfo{author}{\bibfnamefont{T.}~\bibnamefont{Steiner}},
  \bibinfo{author}{\bibfnamefont{M.~L.~W.} \bibnamefont{Thewalt}},
  \bibinfo{author}{\bibfnamefont{E.~S.} \bibnamefont{Koteles}},
  \bibinfo{author}{\bibfnamefont{J.~Y.} \bibnamefont{Chi}}, \bibnamefont{and}
  \bibinfo{author}{\bibfnamefont{B.}~\bibnamefont{Elman}},
  \bibinfo{journal}{Phys. Rev. B Rapid Comms.} \textbf{\bibinfo{volume}{38}},
  \bibinfo{pages}{R3583} (\bibinfo{year}{1988}).

\bibitem[{\citenamefont{Schmidt and Ulbrich}(1996)}]{donorRRS}
\bibinfo{author}{\bibfnamefont{S.}~\bibnamefont{Schmidt}} \bibnamefont{and}
  \bibinfo{author}{\bibfnamefont{R.}~\bibnamefont{Ulbrich}},
  \bibinfo{journal}{23rd Int. Conf. on the Phys. of Semiconductors, Vol. 4} pp.
  \bibinfo{pages}{2741--2744} (\bibinfo{year}{1996}).

\end{thebibliography}

\end{document}